\documentclass[superscriptaddress,reprint,amsmath,amssymb,aps,
floatfix,
]{revtex4-1}
  
\usepackage{graphicx}
\usepackage{dcolumn}
\usepackage{bm,MnSymbol}
\usepackage{units}
\usepackage{psfrag}
\usepackage{float}
\usepackage{color}

\newcommand{\ket}[1]{\vert #1 \rangle}

\begin{document}

\title{Entanglement-based wavelength multiplexed quantum communication network}
\thanks{Correspondence and requests for materials should be addressed to
 S\"oren Wengerowsky and Rupert Ursin.}

\author{S\"oren Wengerowsky}
\email{Soeren.Wengerowsky@oeaw.ac.at}
\affiliation{Institute for Quantum Optics and Quantum Information - Vienna (IQOQI), Austrian Academy of Sciences, Vienna, Austria}
\affiliation{Vienna Center for Quantum Science and Technology (VCQ), Vienna, Austria}
\author{Siddarth Koduru Joshi}
\affiliation{Institute for Quantum Optics and Quantum Information - Vienna (IQOQI), Austrian Academy of Sciences, Vienna, Austria}
\affiliation{Vienna Center for Quantum Science and Technology (VCQ), Vienna, Austria}
\author{Fabian Steinlechner}
\affiliation{Institute for Quantum Optics and Quantum Information - Vienna (IQOQI), Austrian Academy of Sciences, Vienna, Austria}
\affiliation{Vienna Center for Quantum Science and Technology (VCQ), Vienna, Austria}
\author{Hannes H\"ubel}
\affiliation{Optical Quantum Technology, Digital Safety \& Security Department, AIT Austrian Institute of Technology GmbH, Donau-City-Str. 1, 1220 Vienna, Austria}
\author{Rupert Ursin}
\email{Rupert.Ursin@oeaw.ac.at}
\affiliation{Institute for Quantum Optics and Quantum Information - Vienna (IQOQI), Austrian Academy of Sciences, Vienna, Austria}
\affiliation{Vienna Center for Quantum Science and Technology (VCQ), Vienna, Austria}

\date{\today}
\begin{abstract}

Quantum networks scale the advantages of quantum communication protocols to more than just two distant users. Here we present a fully connected quantum network architecture in which a single entangled photon source distributes quantum states to a multitude of users. Our network architecture thus minimizes the resources required of each user without sacrificing security or functionality. As no adaptations of the source are required to add users, the network can readily be scaled to a large number of clients, whereby no trust in the provider of the quantum source is required. Unlike previous attempts at multi-user networks, which have been based on active components, and thus limited to some duty cycle, our implementation is fully passive and thus provides the potential for unprecedented quantum communication speeds.
We experimentally demonstrate the feasibility of our approach using a single source of bi-partite polarization entanglement which is multiplexed into 12 wavelength channels to distribute 6 states between 4 users in a \emph{fully  connected graph} using only 1 fiber and polarization analysis module per user.

\end{abstract}

\maketitle

\section{\label{sec:Quantum}Quantum key distribution Networks}
Quantum Key Distribution (QKD)~\cite{gisin2002,scarani2009security} has reached the level of maturity required for deployment in real-world scenarios~\cite{Stucki2011,Sasaki2011,Peev2009,Xu2009,Elliott2005}, and has been shown to operate alongside classical communication in the same deployed telecommunication fiber~\cite{Choi2011,mao2017integration,patel2012coexistence} and even over long distances in both fiber~\cite{Korzh2014,yin2016measurement} and free-space links~\cite{Ursin2007,bourgoin2015free,liao2017satellite,takenaka2017satellite,gunthner2017quantum}.

Despite these great advances, the practical applicability of QKD is severely curtailed by the fact that most implementations and protocols are limited to two communicating parties.

The pressing need to adapt quantum communication to more than two users has motivated several attempts at quantum networks. The QKD networks demonstrated thus far can be roughly grouped into four types of configurations~\cite{Yang2017networks}: 

First, Quantum repeater networks~\cite{kimble2008quantum} which  use quantum memories and entanglement swapping to extend and route quantum states and form arbitrary network topologies. However, technological advancement in quantum memories are needed until quantum repeater networks can be considered practical. Note that quantum  repeaters can also be used to improve the performance of the following types of quantum networks. 

Another approach to multi-user networks is to use high-dimensional/multi-partite entanglement to share entanglement resources between several users~\cite{Ma2007a,pivoluska2017layered, Torma1999}. This way different users share different subspaces of the Hilbert space to generate their keys. However, adding or removing users requires changes in the dimensionality of the system which makes complex alterations of the source necessary.

The third approach are trusted node networks: They amount to a mesh of point-to-point links, each requiring a complete two-party communication setup. While trusted nodes have been used to extend  bipartite quantum communication schemes to larger multi-user networks~\cite{Peev2009,Sasaki2011,Ribordy1998,Stucki2011,Xu2009}, they also  relinquish the strong security offered by quantum cryptography. Furthermore, this approach creates a significant resource overhead, as it duplicates sender and receiver hardware. 

The fourth approach is to realize a point-to-multipoint network consisting of two (or more) sets of users, whereby a member of the first set can communicate with any member of the second set, but not with members of the same set. This type of configuration allows multiple users to share receivers or sources and has been realized in configurations with passive beam splitters~\cite{Townsend1997,Choi2011,Frohlich2013}, active transparent switches which establish a temporary quantum channel between two particular users at a time ~\cite{Toliver2003,Chen:10-chinese-access-network,Elliott2005,Chang2016}, and frequency multiplexing  ~\cite{Aktas2016,Chang2016,Zhu15,lim2008}.

Fully connected implementations have, in principle, been demonstrated using an entangled photon source~\cite{Herbauts13}. There the bi-partide state was actively routed from one pair of users to any arbitrary other pair of user.

Here we present a network architecture and its realization as a fully connected network without  any active switching which does not limit the distribution rate as per the duty cycle of the switching device anymore.

We connect 4 users to a polarization-entangled photon source via a single optical fiber each, and use the frequency correlations of the photons in combination with wavelength division multiplexing (WDM) to distribute bi-partite entanglement between all pairs of users. 

This allows all pairs of users to generate their own private key while using only a single source.

In addition to the technical advantages of a fully passive implementation, in particular when operating at higher repetition rates, our approach exhibits additional benefits when adding users, as well as in terms of required resources and scalability to larger number of users.

The transition to all-passive optical networks offers a significant boost in terms of reliability, speed and miniaturization. Further, a commercially viable network enables easy modification of the network topology or adding/removing of users without  affecting the key rates of all existing users.

\section{\label{sec:archi}Network architecture}
The fully connected graph is the most robust and flexible network architecture. It  can readily be adapted to any other topology.
Allowing each pair of users to communicate, effectively creates a logical topology akin to a fully connected graph. 

\begin{figure}
    \centering

\includegraphics[width=0.5\textwidth]{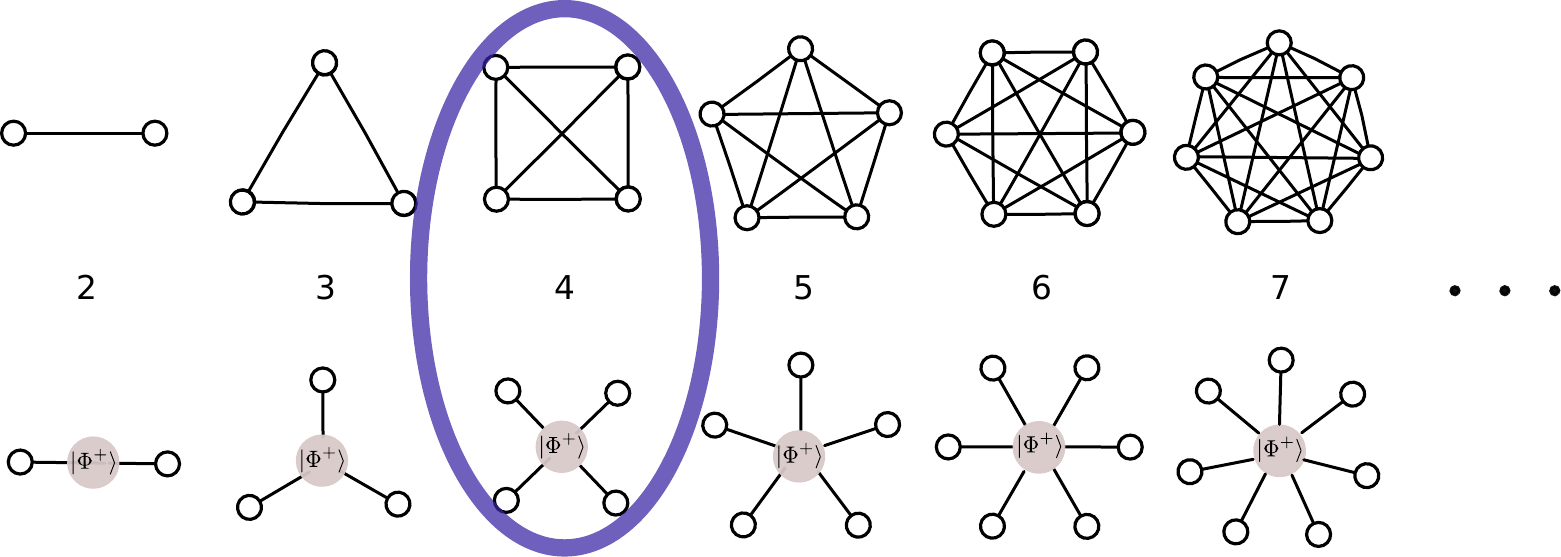}
    \caption{Possible configurations using a complete graph as network for 2 to 7 users. The upper row shows the quantum communications links, while the lower one shows the necessary physical fiber connections. We implemented the case of 4 users. }
    \label{fig:mandala}
\end{figure}

The upper row of Fig.~\ref{fig:mandala} shows the logical topology for up to 7 users with the circles representing users and the lines representing logical links. In the past, each of these links were implemented using separate two-party quantum communication setups~\cite{Peev2009,Sasaki2011,Ribordy1998}. Our approach uses a single source of bipartite entanglement combined with auxiliary correlations in another degree of freedom to realize our entire network, represented by the grey circles in the lower half of Fig.~\ref{fig:mandala}. Each of the $N$ users is equipped with a single detection module, exactly the same as in standard two-party quantum communication schemes. One single mode fiber per user enables communication between each user and every other user. The service provider (i.e., owner of the source of entanglement) multiplexes $N-1$ channels into each single mode fiber. Thus, using $N(N-1)$ channels,  entangled photon pairs (and hence secure keys) can be shared by any two users because the emitted photon pairs are distributed among the channels. As the number of users increases, the resulting network's physical topology remains elementary and grows linearly as shown in the bottom row of Fig.~\ref{fig:mandala}. However, the logical topology becomes increasingly complex and grows quadratically. This allows us to easily create large complex networks. 

Our experimental demonstration of the network architecture exploited WDM and a polarization entangled photon pair source. These choices are specific to the implementation and are not fundamental to the network architecture/logical topology. Our architecture can also be implemented using Time Division Multiplexing (TDM) and or time-bin entanglement instead.

\begin{figure}[ht]
    \centering
    \includegraphics[width=0.5\textwidth]{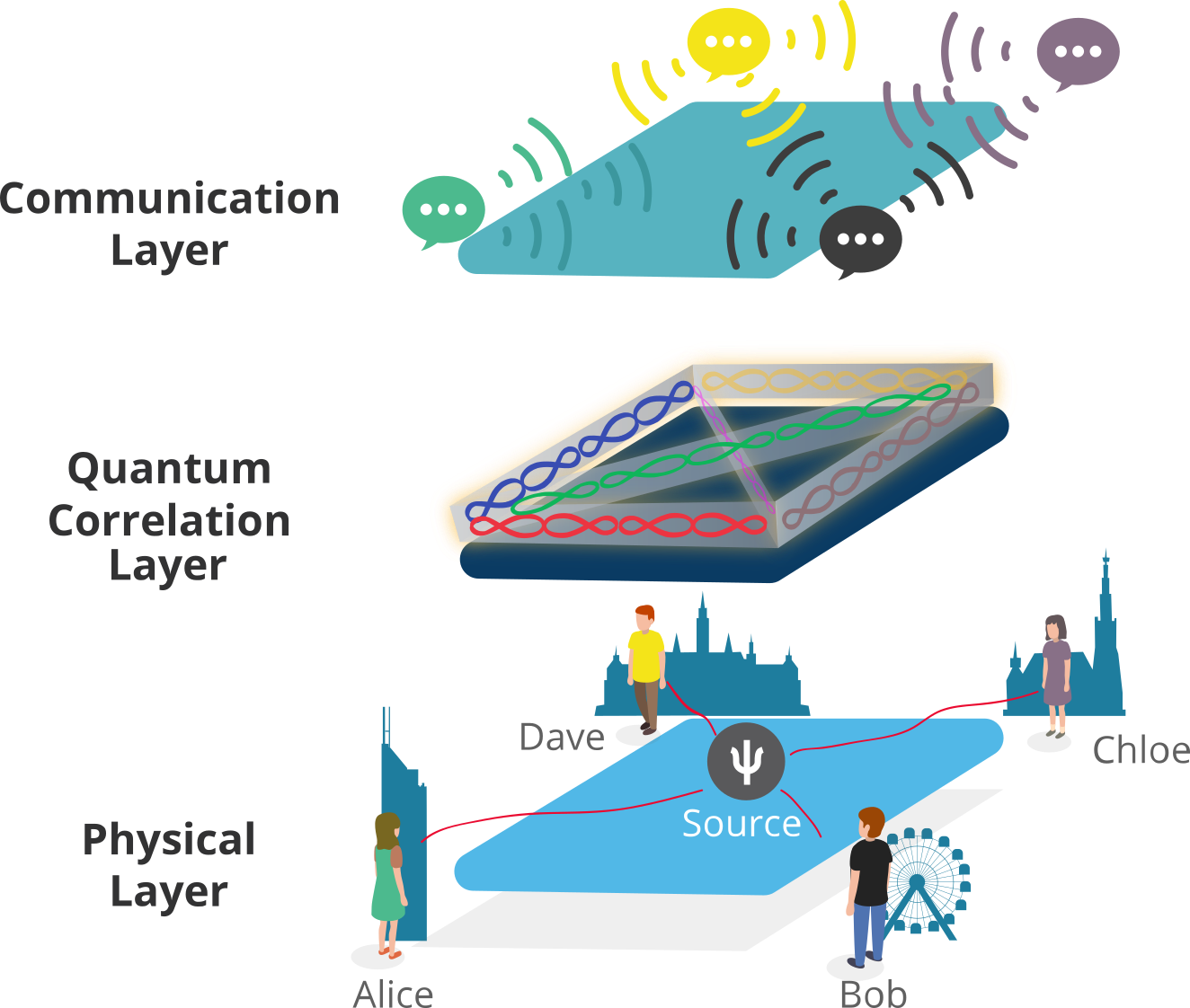}
    \caption{\small Our Network architecture: Different layers represent different levels of abstraction. \emph{Physical connections layer:} contains all tangible components. Each of the 4 users receives a combination of 3 wavelength channels via a  single mode fiber. Thus, the source distributes 6 bipartite entangled photon states to the four users Alice, Bob, Chloe and Dave. \emph{Entanglement distribution layer:} shows the 6 entangled states (each corresponding to a different secure key) that link the 4 users. \emph{Communications Layer:} Entanglement-based two-party QKD protocols like E91 can be used to generate secure keys between all pairs of users. }
    \label{fig:layers}
\end{figure}

We implemented our network architecture with 4  simultaneously active users. The complete network architecture is shown in Fig.~\ref{fig:layers} and can be better understood if divided into layers of abstraction. The bottom layer represents the physical connections/topology. The middle layer represents the shared entangled states. The upper layer represents the classical communication, post processing and secret keys (black arrows) exchanged between users.

 Scalability of the network is of utmost importance, i.e. adding users to the network should not increase its complexity. To add a new user into a network which uses our architecture, the service provider must simply multiplex more channels into each user's fiber. No changes need be made to the source of entanglement, the type of quantum state produced, the user's hardware or even to the software/classical post processing. In any case, only  a detector/receiver unit is required on the user side. 

The architecture does not limit the topology  to a fully-connected graph (as shown in Fig.~\ref{fig:mandala}) but can be adapted  to all possible sub-graphs.

\section{\label{sec:Setup}Setup}

\begin{figure*}
    \centering
    \includegraphics[width=0.75\textwidth]{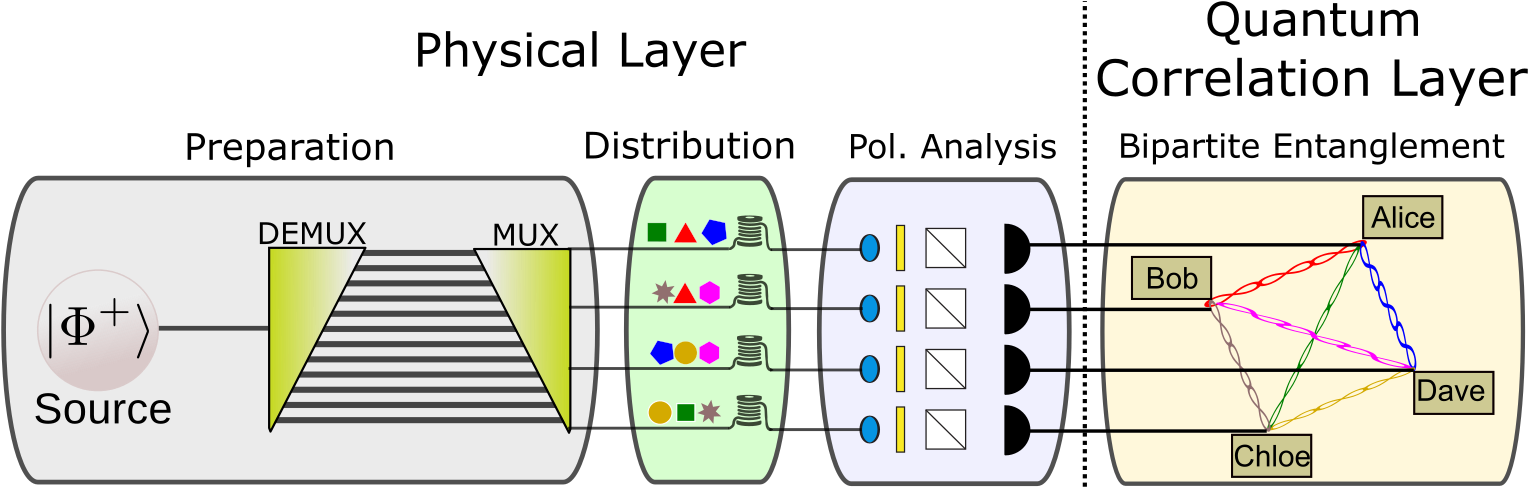}
    \caption{Experimental scheme: 12 wavelength channels are distributed to four users  and therefore allow to establish 6 pair-wise entangled connections. The pairs of symbols (red triangles, green squares, etc.) indicate the pairs of photons.}
    \label{fig:schematics}
\end{figure*}

 Figure~\ref{fig:schematics} depicts an overview of the experimental setup. Photon pairs from a polarization entangled source are separated into different wavelength channels. Specific channels are multiplexed into a single fiber and therefore passively rerouted to each user. 

\subsection{\label{sec:Methods}Entangled photon source} 
To implement our network architecture based on commercially available Dense WDMs (DWDMs), we developed a novel source of frequency correlated polarization-entangled photon pairs at telecommunications wavelengths.

The source was based on type-0 spontaneous parametric down-conversion in a 4-cm-long Magnesium Oxide doped periodically poled Lithium Niobate (MgO:ppLN) bulk crystal with a poling period of 19.2 $\mu$m. The type-0 process converts, with a low probability, one pump photon \unit[775.075]{nm} from a CW Laser to co-polarized signal and idler photons in the telecom C-band.

\begin{figure*}
\centering
   \hspace{-15ex}  \includegraphics[width=0.9\textwidth]{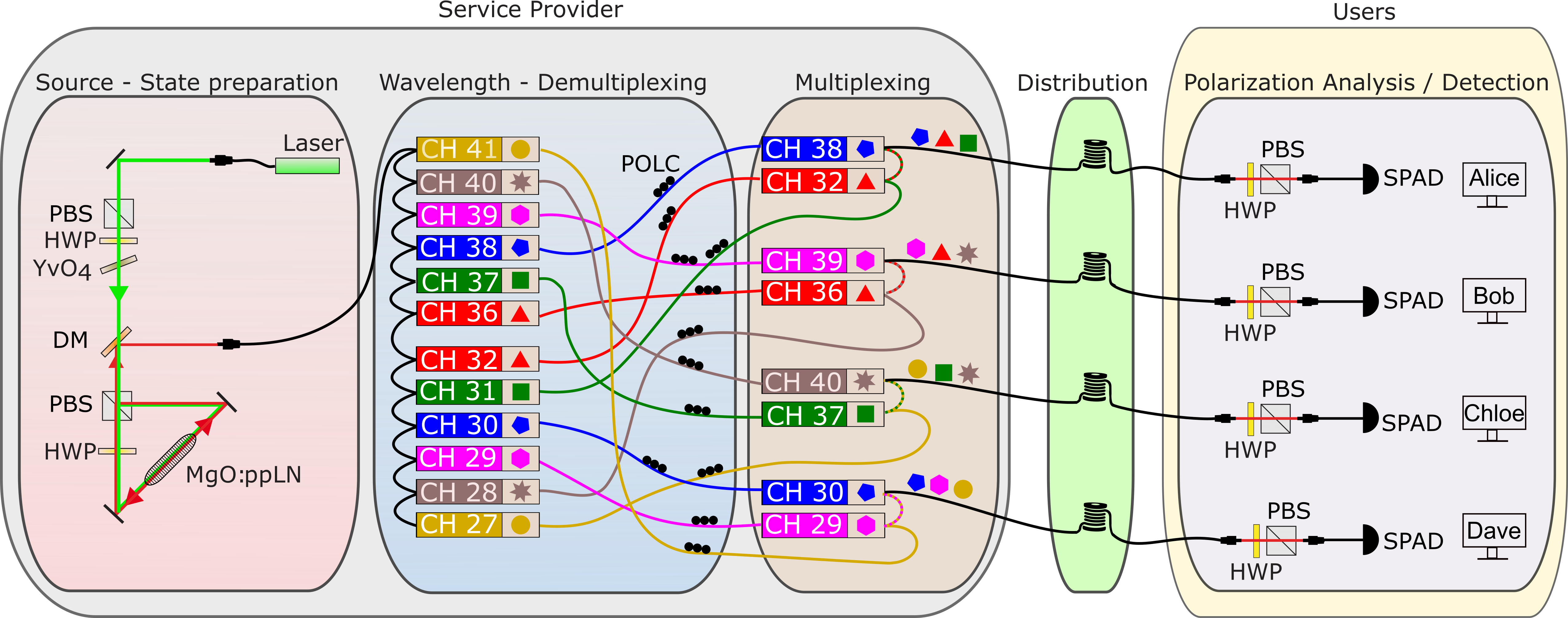}
    \caption{Setup: A Laser at 775 nm (green beam) is used to pump a temperature stabilized MgO:ppLN crystal in a Sagnac-type configuration to create a polarization entangled state. The spectrum is split afterwards into a cascade of 12 ITU channels using band-pass filters. The resulting 12 frequency channels were combined into 4 single mode fibers such that each user (Alice, Bob, Chloe, Dave) receives 3 frequency channels and therefore shares a polarization-entangled pair with each of the other nodes. Abbreviations: DM: dichroic mirror, HWP: half-wave plate, PBS: polarizing beams-splitter, MgO:ppLN, MgO-doped periodically poled lithium niobate crystal, POLC: manual polarization controllers, SPAD: single photon avalanche diode detectors, YVO$_4$:\ yttrium orthovanadate plate. }
    \label{fig:source+wdms}
\end{figure*}

The MgO:ppLN crystal was bidirectionally pumped inside a Sagnac-type setup (see Fig.~\ref{fig:source+wdms}) \cite{Kim2005,lim2008}, thus creating a polarization-entangled state in two wavelength channels: 
\begin{equation}\label{eq:state}
|\Phi\rangle =\frac{1}{\sqrt{2}}( |V_{\lambda_1}V_{\lambda_2}\rangle + |H_{\lambda_1}H_{\lambda_2}\rangle)
\end{equation}

\begin{figure}
    \centering

    \includegraphics[width=0.5\textwidth]{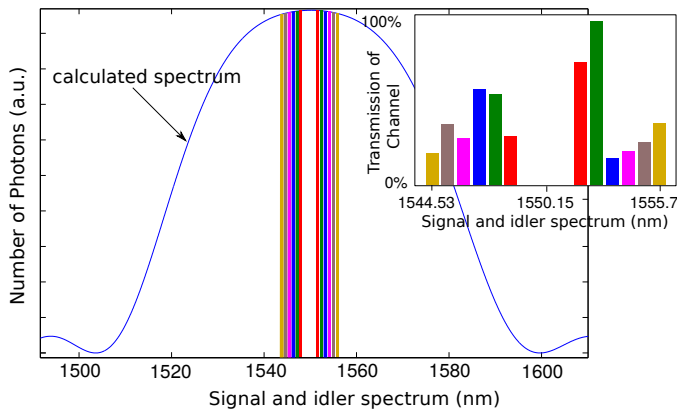}
    \caption{Spectrum for signal and idler. The colourful bars represent the band-pass filters used. The colours indicate the entangled photon pairs. The spectrum of the source was calculated using Sellmeier equations for the MgO:ppLN crystal used in the source~\cite{gayer2008sellmeier} and information about the periodic poling from the supplier.}
    \label{fig:spectrum}
\end{figure}

The spatial mode containing the signal and idler photons from the source was coupled into one single mode fiber and spectrally split by a cascade of band-pass filters. The spectrum of the signal and idler photons was centered at \unit[1550.15]{nm} (see Fig.~\ref{fig:spectrum}) and the filters were chosen to be symmetric w.r.t. this center wavelength. We used \unit[100]{GHz} band-pass filters as defined by the International Telecommunication Union (ITU) in G.694.1.  On the red side of the spectrum we used ITU frequency channels 27-32 and channels 36-41 on the blue side. Due to the well defined pump wavelength of the CW laser and energy conservation during down-conversion, we obtained polarization entanglement between pairs of channels (27 \& 41, 28 \& 40, and so on).  Each user receives 3 channels (see Figure~\ref{fig:spectrum}) via one fiber and used a polarization analysis module to measure in the HV or DA polarization basis.  Single-photon detection events were time-tagged and two-photon  coincidence events were identified within a coincidence window of 1 ns. Fiber polarization controllers were used to neutralize the birefringence of the optical fibers.

Ultimately, the source distributed 6 pairs of polarization-entangled photon pairs between 4 different users successively in a way that each  pair of users shares one pair of photons with each other. 

\subsection{\label{sec:WDM}De-multiplexing and Multiplexing}

The energy correlations of the signal and idler photons (see Eq. \ref{eq:state} and  Fig.~\ref{fig:spectrum}) produced by the source were exploited to separate these modes into separate fibers (De-multiplexing). 
The channels that we used for the signal photons were 27 to 32, while the idler photons were collected in the channels 36 to 41. The corresponding wavelengths can be seen in table \ref{tab:fidelities}.  Entangled photon pairs are found in pairs of channels that have the same spectral distance from the center wavelength. That means, that the channel pairs 27 and 41, 28 and 40, and so on, each share a polarization entangled state (see Fig.~\ref{fig:spectrum}).

The 12 channels have been combined into four fibers using two band-pass filters per fiber, so that three channels reach each one of the 4 users via one fiber. This way, every pair of users shares a pair of channels and therefore entangled photons (see Fig.~\ref{fig:source+wdms} and Fig.~\ref{fig:schematics}).

\section{Results}\label{sec:results}
\subsection{Individual Fidelities}
To characterize the performance of the entangled photon source we measured the fidelity of the state produced as compared to a $\ket{\phi^+}$ Bell state. This measurement was performed directly after de-multiplexing (i.e., splitting of the signal and idler photons) but before the multiplexing of several channels to each user. 
For this measurement, the polarization entanglement visibility was measured in all 3 mutually unbiased bases just after the first cascade of band-pass filters. In this case, the fibers were only compensated in one basis (HV) and the pump state of the source was changed for each measurement in order to compensate for the other basis. It was important to confirm that the source can provide high quality entanglement in all available channel pairs. 

\begin{table}
    \begin{tabular}[b]{c|c|c}
ITU Ch. Numbers & Channel Wavelengths (nm) &Fidelity  ($\pm 0.3 \%$)\\
\hline 
 27 / 41 & 1555.75 / 1544.53 & 98.0 \% \\
 28 / 40 & 1554.94 / 1545.32  &   98.7 \% \\
 29 / 39 & 1554.13 / 1546.12  & 99.1 \% \\
 30 / 38 & 1553.33 / 1546.92 & 99.0 \% \\
 31 / 37 & 1552.52 / 1547.72 & 99.2  \% \\
 32 / 36 & 1551.72 / 1548.52   & 97.3  \% \\              
\end{tabular}
    \caption{Bell-State-Fidelity of the entangled state produced by the source directly measured at the channel pairs before multiplexing.}
    \label{tab:fidelities}
\end{table}

\subsection{Fidelities with Multiplexing}
Once we confirmed that the source of entangled photon pairs was able to provide  high quality entanglement we connected the multiplexers and sent 3 channels to each of the 4 users. 

\begin{figure}[htb]
    \centering
    \includegraphics[width=0.5\textwidth]{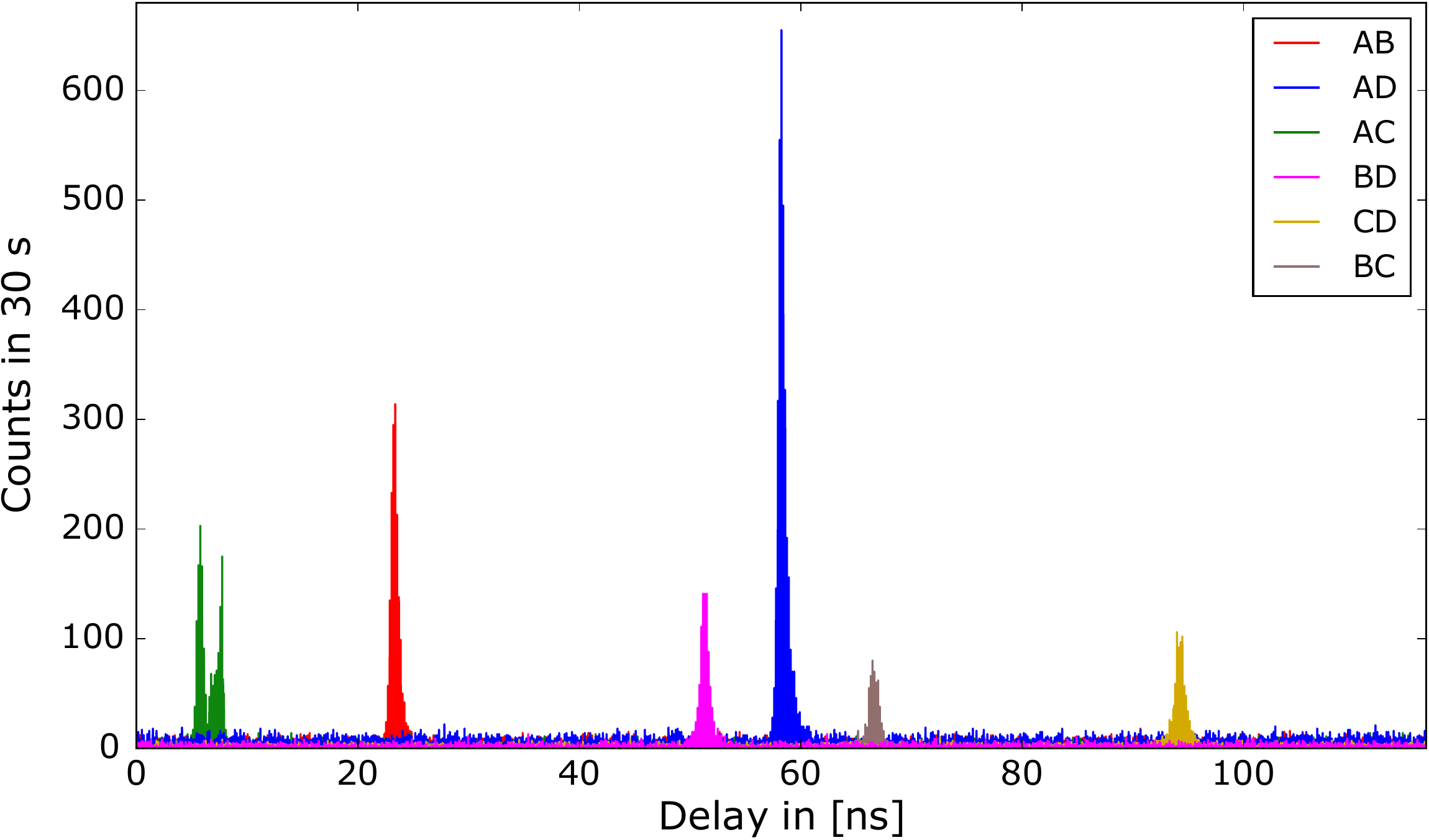}
    \caption{Temporal cross correlations between the time-traces of the 4 detectors. Each one  shows a distinct peak. This way, the coincidence clicks were identified.}
    \label{fig:g2}
\end{figure}

To measure the fidelities, all 12 fiber channels had been  compensated in two mutually unbiased bases from the source until the measurement module in order to demonstrate that the entangled states are  simultaneously created in all channels without further alignment.  Further, the multiplexing was implemented, so that three channels were detected on each of the four detectors.  Entangled pairs were identified using the temporal cross-correlation functions (see Fig.~\ref{fig:g2}).

We used 4 free-running single photon avalanche detectors based on a passively quenched InGaAs avalanche photodiode. Three detectors operated at a detection efficiency of 2-3\% and a dark count rate between 350 and 1500\,Hz with a dead time of 1 $\mu$s for the measurement modules ``Bob'', ``Chloe'' and ``Dave''. The measurement module ``Alice'' employed a detector with about 10\,\% efficiency, 1000\,Hz dark counts and 4\,$\mu$s dead time. The coincidence rate differed between 10 and 65\,Hz for the six entangled links as the losses and detection efficiencies were unequal. We measured the visibility of all 6 entangled links in 2 mutually unbiased bases HV and DA and computed the fidelity.  This amounts to 16 different basis settings for the HV basis as well as for the DA basis. Each basis setting was measured for 30 seconds. The singles rates of the four detectors were between 21\,kHz and 73\,kHz. An overview over the raw counts for the setting HHHH, where the maximal coincidence rate is expected, is given in table \ref{tab:counts}.

\begin{table}
    \begin{tabular}[b]{c|c|c|c|c}
 & Alice & Bob & Chloe & Dave\\\hline
Alice & 2204203 & 2049 & 1156 & 3813\\
Bob & & 878692 & 569 & 1018 \\
Chloe & & & 636268 & 748\\
Dave & & & & 1231478 \\
\end{tabular}
    \caption{Measured Coincidence- and Single counts in 30 seconds for all four measurement stations at the setting HHHH.}
    \label{tab:counts}
\end{table}

\begin{figure}[htb]
    \centering
    \includegraphics[width=0.5\textwidth]{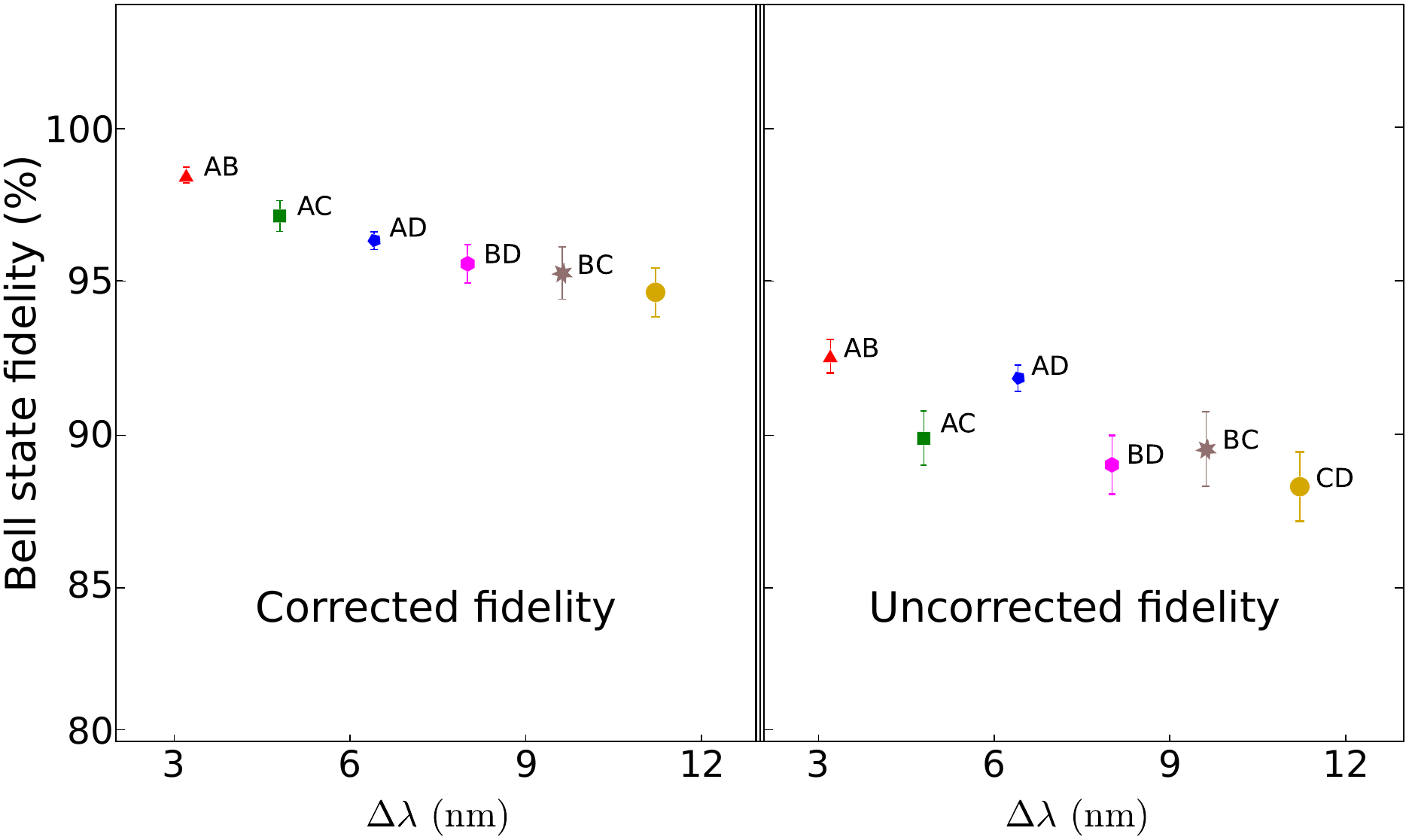}
    \caption{Measured fidelities  with and without subtraction of accidental coincidences. Each point is measured using two WDM channels which connect the respective users. The X-axis represents the difference in wavelength between the respective channels of the respective two partner photons.}
    \label{fig:visi}
\end{figure}

Figure~\ref{fig:visi} shows the results of the Bell state fidelity measurements. Due to the timing uncertainty of the detectors, we are limited to a rather large coincidence windows of 1\,ns. This way, detector clicks are falsely identified as pairs and deteriorate the measured fidelity. The right hand side of the figure shows the fidelity corrected for this error while the uncorrected values are shown to the left.

Using the uncorrected fidelities (see Fig. \ref{fig:visi}) and count rates, we estimated a raw key rate between 10 and 34\,Hz, which would yield a secure key rate between 3 and 15\,Hz \cite{Ma2007}.
A fidelity larger than 81 \% is necessary to obtain a positive secure key rate. Therefore, using their polarization detection modules, the users were able to measure a non-classical polarization correlation visibility in the HV and DA bases from which we can calculate the lower bound on the Bell state fidelity. These measurements  show that we have successfully shared an entangled state between every pair of users.

\section{Conclusions}

We have successfully realized a proof of principle demonstration of quantum communication network. The network architecture can be readily adapted to any other network topology and additional users can be added with minimal modifications.

This network architecture allows us to consider practical usage scenarios very similar to existing everyday networks (see section~\ref{use} of the Supplementary Information). It is conceivable that this networking concept can be combined with other access networking ideas as proposed in \cite{Ciurana14}. 

We have implemented this network at telecommunication wavelength and it is thus compatible with existing infrastructure.
The scalability and ease of upgrading of this network architecture make it one of the best candidates for commercial quantum communication networks. 

\subsection{Discussion}\label{sec:conc}

The network continues to offer all the security benefits of entanglement based QKD and does not require trusted nodes. In contrast to networks based on active switching~\cite{Herbauts13,Chang2016,Chen:10-chinese-access-network}, the only limit on the communication speed in our passive scheme is given by the brightness of the source and the ``quality'' of the detector (efficiency, timing jitter and dead time).  The finite duty cycle and switching rate of a possible active component do not limit our network.

Minimizing the resource requirements while maintaining full connectivity is thus a key requirement. 

Naively, an alternative method to implement a fully connected quantum network with a similar topology would be to use a 1:$N$ beam-splitter and probabilistically distribute entangled photon pairs between all users. 
 The main benefit of our wavelength multiplexed implementation reveals itself when each user opts to de-multiplex the different wavelength channels onto $m$ single-photon detectors (where $1<m<N$) In this case, due to the deterministic frequency correlations, every pair of frequency channels can be considered an independent communication link and a $m$-fold increase in the total key generation rate is achieved while maintaining the same signal-to-noise ratio of a two-party communication. Conversely, probabilisitc distribution using a 1:$N$ beam-splitter would always reduce the signal-to-noise ratio as users are added. 
 
Our network is linearly scalable in terms of client resources and additional users can be added to the network without changing client hardware (see section~\ref{scale} of the Supplementary Information.)

\subsection{Outlook}

An interesting question to ask is, how many users can be added to this network architecture while maintaining its performance. In our implementation, we used one detector per user to detect all three frequency channels.
As mentioned above, compared to a two-party communication scheme, detecting more than one channel on the same detector gives a higher noise level because the singles count rate is tripled and the coincidence rate per link is unchanged. The measured fidelities show that the network architecture is sound despite the increased level of noise. 
The number of available wavelength channels within the entangled photon spectrum and the performance of the detectors used (dark counts, timing jitter and efficiency) also contribute to the limit on the number of users. For a more detailed discussion of the expandability of our network, see the supplementary information section~\ref{SNR}).

However, this limitation can be avoided, because all the users have the option to split the signal to detect only a few or just one frequency channel per detector in order to recover the signal-to-noise ratio of a two-party communication. Alternatively, groups of users could temporally block frequency channels which are not currently needed for their communication. 
This way, the network could also be used like an access-network with switching on the client-side. 

Instead of a continuous entanglement distribution, it is also conceivable to  use a pulsed pump laser for the entangled photon source as described in section~\ref{pulse} of the Supplementary Information. This would improve the signal to noise-ratio because it would allow to detect the communication with different users in different time slots, similar as in~\cite{Zhu15}.

Apart from standard entanglement-based QKD protocols, also distributed computation tasks like the millionaire's problem~\cite{HE2013} as well as Byzantine fault tolerance and asynchronous reference frame agreement~\cite{islam2016asynchronous} can be implemented on this network.

\subsection{Acknowledgements}

Financial support from the Austrian Research Promotion Agency (FFG) -Agentur für Luft- und Raumfahrt (FFG-ALR contract 844360 and FFG/ASAP:6238191 / 854022), the European Space Agency (ESA contract 4000112591/14/NL/US), the Austrian Science Fund (FWF) through (P24621-N27) and the START project (Y879-N27), as well as the Austrian Academy of Sciences is gratefully acknowledged. We would also like to thank Jesse Slim for his help with the software and Evelyn Ortega for assistance in the lab.

\appendix

\section*{Supplementary Information}\label{appendix}
\subsection{QKD and signal-to-noise ratio considerations}\label{SNR}
\begin{figure*}
    \centering
    \begin{minipage}{.5\textwidth}
    \includegraphics[width=0.95\textwidth]{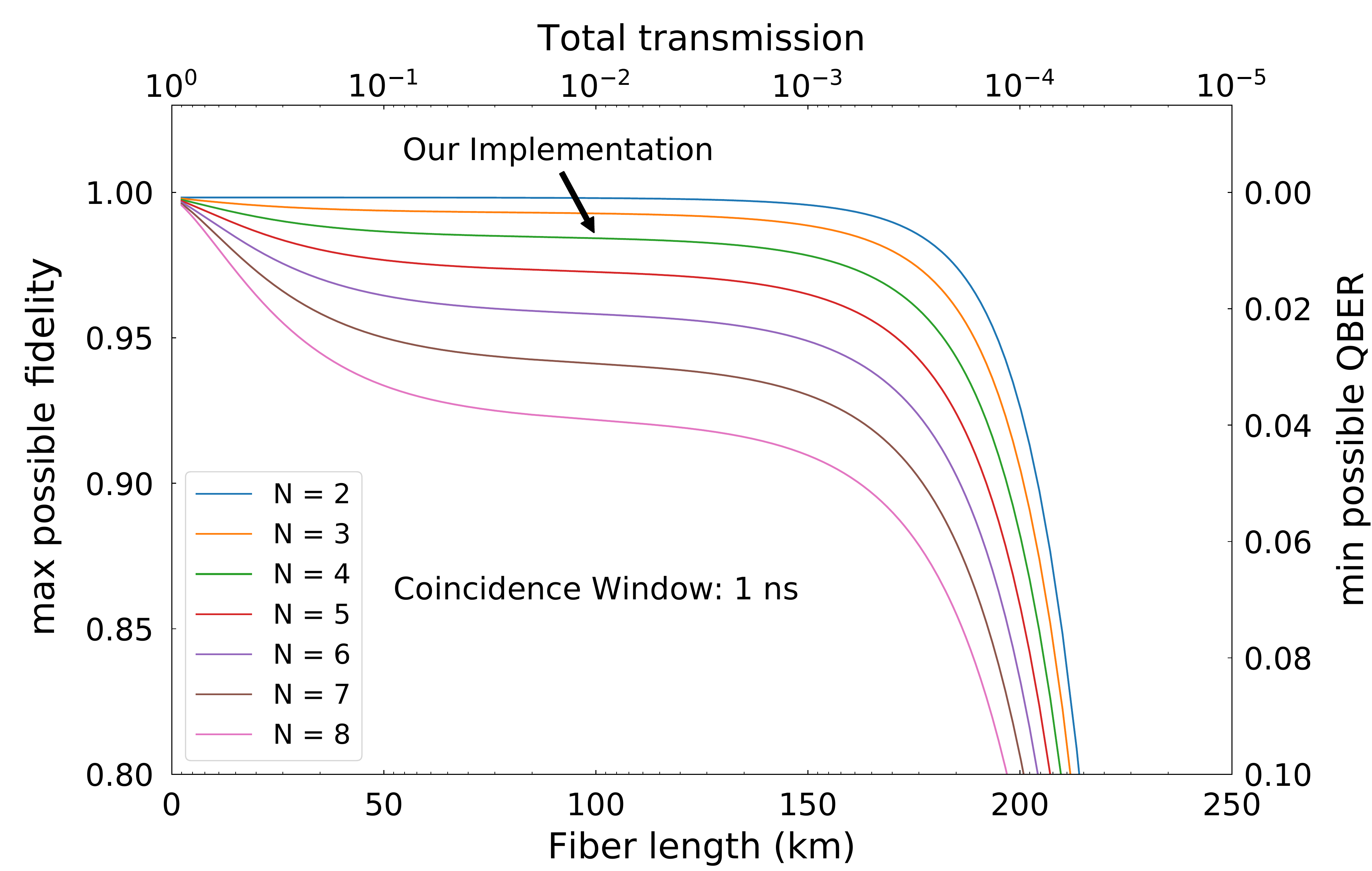}
    \end{minipage}%
    \begin{minipage}{.5\textwidth}
    \includegraphics[width=0.95\textwidth]{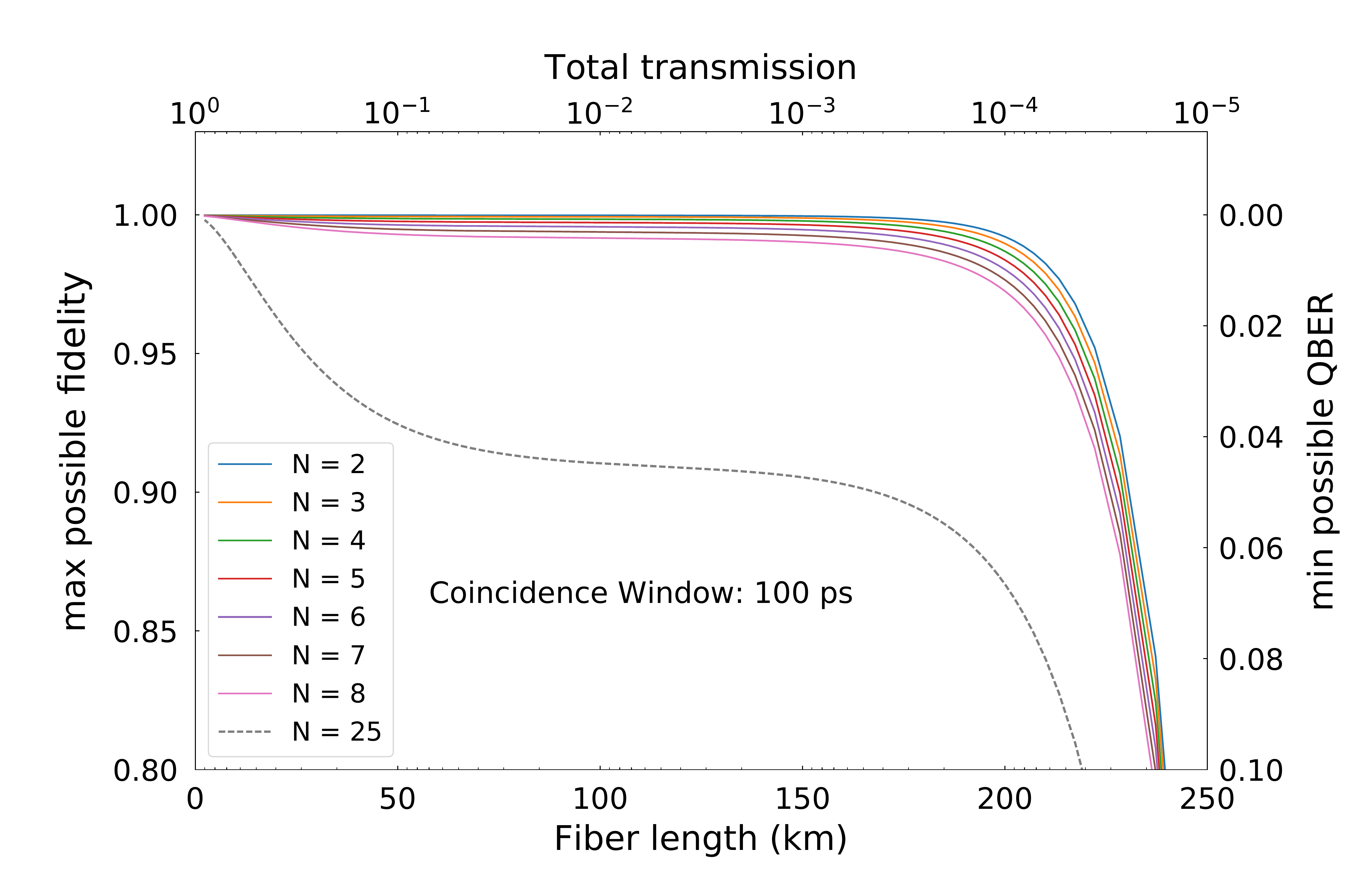}
    \end{minipage}
    \caption{Calculated fidelities / QBERs for 2 to 9 users vs system efficiency. \textbf{Left}: Using detectors with a 1\,ns timing jitter. This is great for cheap networks with low losses (i.e., over a small area like a LAN). \textbf{Right}: Using detectors with a 100\,ps jitter allows us to sustain much higher losses and many more users. This is useful for long distance intercity links. Both graphs shown above were calculated using a generated pair rate of $1.7$ million pairs/s and a dark count rate of 500/s per detector. 
    \label{fig:numnodes}}
\end{figure*}

In general the chief drawback of this architecture is the amount of noise introduced by detecting several channels on a single detector. The noise results in a loss of fidelity (or quality of the entanglement). 
In order to implement a QKD scheme, all users would announce their time-tags and correlation functions (see fig. \ref{fig:g2}) publicly, so that everybody is able to ignore counts that don't belong to their communication and therefore improve the signal-to-noise ratio. This improvement is related to the total losses in the system as shown in Fig~\ref{fig:numnodes}. This is equivalent to ignoring all global ($n>2$)-fold events.

In other words, the count rate $S_i$ per user in a network with $N$ nodes,  link and system efficiency $\eta$  and dark count rate $D$ would be reduced by a term that scales proportional to the coincidence probability: 
$$ S_i = D+(N-1)\frac{P}{2}\eta-(N-2)\frac{P}{2} \eta^2$$ 
With $P$ being the total number of available pairs in the spatial and spectral collection mode of the source.
The rate of coincidence clicks can be estimated as follows: 
$$ C = \frac{P}{2}\eta^2+\tau S_i^2 $$ 
The accidental coincidences account for the minimum number of coincidences observable and therefore reduce the contrast: 
$Acc = \tau S_i^2$.

Another significant improvement can be made by decreasing the coincidence time window. This can be achieved by using faster detectors with a much smaller timing jitter. For example reducing the coincidence window to 100\,ps results in a maximum fidelity shown in Fig~\ref{fig:numnodes}.

\subsection{Scalability}\label{scale}
Our network architecture is easily scalable and users can be added and removed easily without any change to the user's hardware. However, like most existing network hardware there are limitations to the scalability. The three main limitations are: First, the brightness of the source which can be overcome by using more or longer waveguides/crystals and stronger pumping. Second, the limited bandwidth of the source dictates how many wavelength channels can be used, nevertheless, this too can be overcome by using narrower wavelength channels. Third, accidental coincidences contribute significantly to the wavelength and increase dramatically with the number of users. This can be mitigated by using the method described in Subsection~\ref{SNR} to help reducing the noise.  Naturally, using faster detectors and therefore shorter coincidence windows can also help to improve this.
A pulsed pump experiment would further mitigate the problem of accidental coincidences by defining fixed time slots for the arrival of each channel at the detector. 

Our network architecture offers the advantage of simultaneous communication between one node and every other node. Nevertheless, should one user choose to completely block the signal from a set of other users, an active switch capable of selecting certain channels can be used. This would allow users to control the network topology and create custom sub-graphs without the intervention of the service provider. Further, detecting only a chosen subset of channels will limit the accidental coincidence rates and allow for faster communication with a chosen sub-graph.

\subsection{Pulsed network scheme}\label{pulse}

The drawback of the scheme presented here is the increase in the accidental count rates due to the multiplexing of many quantum channels onto a single detector. This limitation can be completely overcome by using a pulsed scheme. Consider the experiment presented here with a pulsed laser where, the pulse width is much smaller than the detector jitter. Further, each of the $n$ users has gated detector(s) for which the gate is opened $n-1$ times for each laser pulse. Each opening of the gate corresponds to the time delay between different coincidence peaks among all users with each user in question. With ideal detectors, the performance of the pulsed scheme will be equivalent to $n-1$ separate quantum communication setups with the same detectors and comparable count rates per link. Interesting, when using real-world detectors like InGaS SPADs which have a significant dead time, the performance of our pulsed network scheme can exceed that of $n-1$ independent setups. When the dead time of the detector is larger than the interval between opening each of the $n-1$ gates/pulse, then a noise count in one gate prevents the occurrence of a noise count in all subsequent gates within the dead time. This suppression of noise clicks can lead to improved key rates and QBER. The advantage is strongest when there is only one photon pair in the given set of $n-1$ links per user.

This pulsed network scheme, would require an additional gating signal to be sent to each user. Further, this pulsed scheme could be unsuitable for mobile nodes because all nodes need to compensate the delays to all other nodes. However, for fixed users, this pulsed network scheme is ideal and can drastically improve the network throughput by reducing the accidental count rate by a factor equal to the duty cycle of the gating.

\subsection{Multiplexing and types of entanglement}\label{types}

The logical network topology we have outlined here is independent of the type of entanglement or multiplexing used. Nevertheless, different types of multiplexing have advantages. For example a scheme based on WDMs has a few advantages over that based on TDMs: First, active switching used in TDM is prone to mechanical breakdown and in more complex networks several switches may need to operate synchronously. Second, A bright source can produce multiple photon pairs within a single coincidence time window. However, the probability that multiple pairs are produced in exactly the same wavelength channels is negligible. Thus WDM based network could have a distinct advantage. Third, Introducing an additional TDM channel will reduce the coincidence rates seen by all users, but an additional WDM channels will not affect already existing connections. Lastly, crosstalk between the  channels is not  harmful, since photons in the wrong channel would, due to the different delay introduced by the WDMs, only contribute to the accidental rate and not be seen as a separate coincidence peak.
On the other hand, a TDM based scheme would need only $2N$ channels. Large scale networks could also combine the advantages of WDMs and TDMs by using both together.

Fiber based quantum communication has often been performed using time-bin entanglement~\cite{Inagaki2013,Aktas2016} to avoid having to compensate for the birefringence of the fiber. However, this requires the service provider and users to have matched and stabilized interferometers (the stabilization of which often requires another stable laser).  Although the logical network topology is compatible with this form of entanglement we chose to use polarization entanglement because it simplifies the user's hardware.
Changes in the birefringence of the optical fiber are easily monitored and compensated for by using regular test signals.

\subsection{Usage scenarios}\label{use}
Quantum communication is unfortunately often thought of as a purely academic concept/experiment. However, the technology is mature enough to consider practical problems about deploying and using QKD links. Typical classical networks consist of smaller Local Area Networks (LANs) and similar as well as much larger inter-city style networks. Both a LAN and an inter-city network have a limited number of users~\footnote{To connect a large number of users together one must interconnect networks to create the ``internet''. Similarly, a quantum internet must also be an interconnection of several networks. A single user in our network architecture could be replaced by a quantum repeater or entanglement swapping setup to interconnect several similar quantum networks.}. The most significant differences between the two types of networks are the distances spanned, the costs and the target market. 

To realize a cheap LAN with current technologies and our new network architecture, we propose using a cheaper type of single photon detector -- ``Single Photon Avalance Diodes (SPADs)''. These typically have a low detection efficiency and large timing jitter. As can be seen by extrapolating  Fig.~\ref{fig:numnodes} (left) the network will be able to tolerate more than 30\,dB of loss with up to 12 users. This loss is more than sufficient to account for a few kilometers optical fiber, the heralding efficiency of a typical source and the poor detection efficiency of SPADs.

Inter-city networks naturally cost far more than a LAN. Our network architecture can be used to build large scale inter-city quantum networks by using high efficiency low timing jitter detectors such as Nano-wire detectors. Figure~\ref{fig:numnodes} (right) shows that we can tolerate more than 43\,dB of loss with up to 25 users spanning distances of $>$ 200\,km.

Further, in any network it is always advantageous to make the user's hardware requirements as simple as possible while the centralized network hardware has the majority of the complexity. We have designed our network architecture along these principles with almost all complexity in the three centralized stages -- source, de-multiplexing and multiplexing.

\end{document}